\documentclass{SciPost}

\hypersetup{
    colorlinks,
    linkcolor={red!50!black},
    citecolor={blue!50!black},
    urlcolor={blue!80!black}
}

\usepackage[bitstream-charter]{mathdesign}
\usepackage[capitalize,noabbrev]{cleveref}
\usepackage{booktabs}
\usepackage{subcaption}
\DeclareSymbolFont{usualmathcal}{OMS}{cmsy}{m}{n}
\DeclareSymbolFontAlphabet{\mathcal}{usualmathcal}

\fancypagestyle{SPstyle}{
\fancyhf{}
\lhead{\colorbox{scipostdeepblue}{\bf \color{white} ~SciPost Physics Proceedings }}
\rhead{{\bf \color{scipostdeepblue} ~Submission }}

\fancyfoot[C]{\textbf{\thepage}}
}

\begin{document}

\pagestyle{SPstyle}

\begin{center}{\Large \textbf{\color{scipostdeepblue}{
Parallel Nested Slice Sampling for Gravitational Wave Parameter Estimation\\
}}}\end{center}

\begin{center}\textbf{
David Yallup\textsuperscript{1$\star$},
Metha Prathaban\textsuperscript{1},
James Alvey\textsuperscript{1} and
Will Handley\textsuperscript{1}
}\end{center}

\begin{center}
{\bf 1} Kavli Institute for Cosmology, University of Cambridge, Cambridge, UK

$\star$ \href{mailto:email1}{\small dy297@cam.ac.uk}\,,\quad
\end{center}

\definecolor{palegray}{gray}{0.95}
\begin{center}
\colorbox{palegray}{
  \begin{tabular}{rr}
  \begin{minipage}{0.37\textwidth}
    \includegraphics[width=60mm]{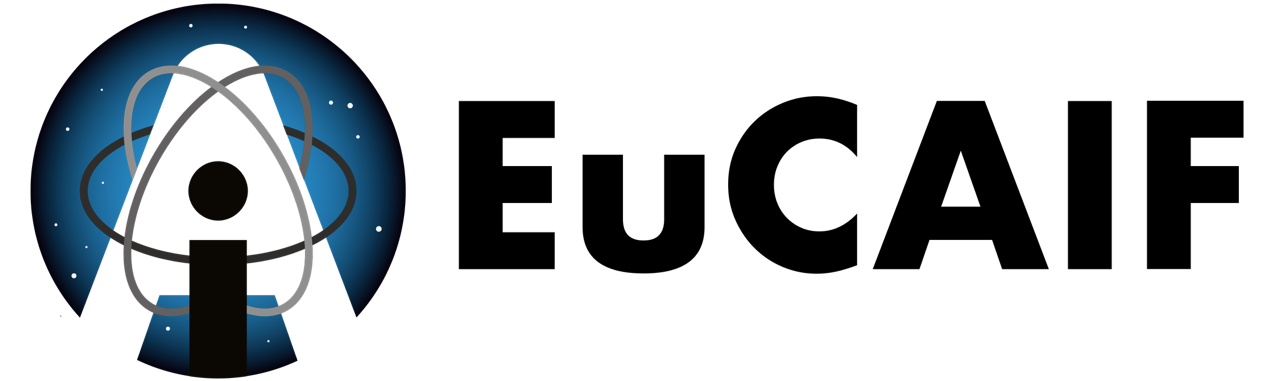}
  \end{minipage}
  &
  \begin{minipage}{0.5\textwidth}
    \vspace{5pt}
    \vspace{0.5\baselineskip} 
    \begin{center} \hspace{5pt}
    {\it The 2nd European AI for Fundamental \\Physics Conference (EuCAIFCon2025)} \\
    {\it Cagliari, Sardinia, 16-20 June 2025
    }
    \vspace{0.5\baselineskip} 
    \vspace{5pt}
    \end{center}
    
  \end{minipage}
\end{tabular}
}
\end{center}

\section*{\color{scipostdeepblue}{Abstract}}
\textbf{\boldmath{%
Inferring parameters and testing hypotheses from gravitational wave signals is a computationally intensive task central to modern astrophysics. Nested sampling, a Bayesian inference technique, has become an established standard for this in the field. However, most common implementations lack the ability to fully utilize modern hardware acceleration. In this work, we demonstrate that when nested sampling is reformulated in a natively vectorized form and run on modern GPU hardware, we can perform inference in a fraction of the time of legacy nested sampling implementations whilst preserving the accuracy and robustness of the method. This scalable, GPU-accelerated approach significantly advances nested sampling for future large-scale gravitational-wave analyses.
}}

\vspace{\baselineskip}

\noindent\textcolor{white!90!black}{%
\fbox{\parbox{0.975\linewidth}{%
\textcolor{white!40!black}{\begin{tabular}{lr}%
  \begin{minipage}{0.6\textwidth}%
    {\small Copyright attribution to authors. \newline
    This work is a submission to SciPost Phys. Proc. \newline
    License information to appear upon publication. \newline
    Publication information to appear upon publication.}
  \end{minipage} & \begin{minipage}{0.4\textwidth}
    {\small Received Date \newline Accepted Date \newline Published Date}%
  \end{minipage}
\end{tabular}}
}}
}




\section{Introduction}

The detection of gravitational waves (GWs) by the LIGO-Virgo-KAGRA collaboration has provided significant advancements in our understanding of the universe, offering new insights into black hole mergers and neutron star coalescences, cosmology, and gravitational theory~\cite{LIGOScientific:2016aoc, LIGO_GWTC1, LIGO_GWTC2, LIGO_GWTC3}. Extracting meaningful information from these signals, however, hinges on robust and efficient inference techniques. Determining the parameters of GW events, such as the masses and spins of the compact objects, and testing competing astrophysical models, often requires computationally intensive Bayesian inference. Nested sampling has emerged as a cornerstone of Bayesian inference in the GW community, providing a powerful framework for both parameter estimation and model comparison. However, despite its robustness and widespread use, nested sampling can be computationally slow, especially when compared to other Markov Chain Monte Carlo (MCMC) methods~\cite{Petrosyan:2022}. This computational bottleneck is a concern, particularly as the volume and complexity of GW data is poised to increase dramatically with next-generation observatories~\cite{Hu:2025}.

To accelerate existing inference tasks and meet the challenges posed by future data, several approaches have been explored. Simulation-Based Inference (SBI) methods, such as neural posterior estimation with implementations like DINGO~\cite{dingo2021, Dax:2022pxd, dingonature}, have demonstrated significant successes in accelerating GW inference and have emerged as a powerful tool in the field. Additionally, efforts have focused on modifying the core nested sampling algorithm, leveraging machine learning tools such as normalizing flows, to accelerate convergence~\cite{Prathaban:2024rmu,Williams:2021qyt}. In this work, we explore a complementary approach: leveraging the parallel processing capabilities of Graphics Processing Units (GPUs) to accelerate nested sampling. While there has been previous work on accelerating MCMC methods on GPUs~\cite{Wong:2022xvh}, we focus on nested sampling. By harnessing the power of modern hardware, we aim to provide an alternative and highly efficient method for GW parameter estimation and model comparison. For current data, this approach can significantly decrease computational demand, enabling the use of a robust and trusted method within the field, but at an accelerated pace.

In this work, we apply a recently developed GPU-accelerated nested sampling framework \cite{yallup2025nested} to the context of GW parameter estimation, complementing the work of Prathaban et al.~\cite{Prathaban:2025qgg}. We focus in this work on demonstrating in a more optimal case, where the likelihood is evaluated on a coarser frequency grid, that we can gain even further computational speedup on real GW parameter estimation problems. We demonstrate that crucially this speedup doesn't just arise from the reduced compute cost of each likelihood call, but the massive parallelism of the core NS algorithm can give dramatic further runtime improvements. This underlines the importance of further developing such accelerated likelihood based inference pipelines for GW inference in the future.


\section{GPU-Accelerated Nested Sampling}

Nested Sampling has become a prominent method for inference on gravitational wave signals. For example, the \texttt{bilby} software~\cite{bilby_paper} (which itself is a central tool in the field) implements nested sampling as one of its core inference algorithms using the \texttt{dynesty} package~\cite{dynesty}. From the optimization perspective, the utilization of HPC CPU hardware is enhanced through process parallelization as implemented in the parallel \texttt{bilby} extension~\cite{Smith:2019ucc}.

Recently, a reformulation of the nested sampling algorithm has been proposed~\cite{yallup2025nested}, and implemented in the \texttt{blackjax} framework~\cite{cabezas2024blackjax}. We use the recommended combination of algorithm choice and settings identified as \emph{Nested Slice Sampling} (NSS) in \cite{yallup2025nested}. This implementation readily integrates with recent developments in GW modeling and inference that also target GPU hardware, namely fast vectorized waveform generation via the \texttt{ripple} package~\cite{Edwards:2023sak} and likelihood evaluation via the \texttt{jim} software~\cite{wong2023fastgravitationalwaveparameter}. A \emph{bilby-like} kernel has been demonstrated for this task using the same GPU NS framework~\cite{Prathaban:2025qgg}. In this work we deploy the default slice sampling based NS kernel (recommended in Yallup et al.~\cite{yallup2025nested}) as a point of comparison. We also focus particularly on a regime that is complementary to the work of~\cite{Prathaban:2025qgg}, when the likelihood is well parallelised by employing likelihood heterodyning~\cite{Cornish:2021lje}. 

In comparison to \texttt{bilby} (\texttt{dynesty}), the \texttt{blackjax} implementation of Nested Slice Sampling (NSS) is similar at a high level: both implement the classic nested sampling algorithm with an MCMC walk to evolve particles~\cite{dynesty, bilby_paper}. In particular, \texttt{bilby} (\texttt{dynesty}) uses a customized random walk proposal, whereas our sampler uses a slice sampling proposal~\cite{nealslicesampling}. The \texttt{blackjax} implementation, however, executes its slice sampling in a vectorized step across the entire population. Combined with a static memory implementation of the particle update, the entire end-to-end algorithm can then run in GPU memory.

We run with nested sampling hyperparameters, relevant to the \texttt{blackjax} implementation, of: a static population of 3000 live points, with short slice sampling chains of 10$\times$ the number of dimensions in length, and we delete half of the live points at each NS iteration. This represents the default recommended values for the number of particles to delete, the length of the short chains is twice what is usually recommended in \texttt{blackjax}, however it is in-keeping with \texttt{bilby} default values on similar problems. We employ a simple default tuning strategy for the slice sampling chains, using the particle covariance to tune direction proposals. This is troublesome for the wrapped phase and polarization angle parameters in particular, hence the large number of repeats to ensure convergence. Providing better tuning that respects the geometry of the parameter space is an area for future work, but we find that the default tuning is sufficient for the data analysis explored in this work. Being able to delete 1500 live points per iteration highlights the impressive capabilities of a GPU-accelerated nested sampling algorithm, probing parallelism that is largely impossible for CPU implementations.

\section{Application to real data}\label{sec:data}

We validate and benchmark our GPU-accelerated nested sampling
pipeline using real gravitational wave data from the GW150914
event, the first direct detection of gravitational waves from a
binary black hole merger~\cite{LIGOScientific:2016aoc}. This analysis allows us to assess the
performance of our implementation, particularly its runtime and the
effective sample size (ESS). For a direct and fair comparison, we compare to the GPU-accelerated MCMC sampler, FlowMC~\cite{Wong:2022xvh}, which is optimized for the same hardware. We note that it has already been shown that FlowMC (steered via the \texttt{jim} package) agree with the results obtained using \texttt{bilby} in this context~\cite{wong2023fastgravitationalwaveparameter}, and it has been shown that \texttt{blackjax} NS can be brought into nearly exact agreement with \texttt{bilby} when deployed with the same inner kernel~\cite{Prathaban:2025qgg}. We follow mostly the default settings of the \texttt{jim} example script included in the code repository for this event. We increase the number of chains from 500 to 1000, probing similar levels of parallelism to the \texttt{blackjax} implementation, as well as increasing reliable convergence. In both cases we exploit the use of likelihood heterodyning~\cite{Cornish:2021lje}. We fix the same reference parameters used to perform the heterodyning between algorithms, and do not include this in the quoted runtimes. We run both algorithms on a single NVIDIA A100 GPU, with 40GB of memory, and a single CPU core. 

We analyze data from the LIGO detectors at Hanford (H1) and
Livingston (L1)~\cite{LIGOScientific:2016aoc}. The IMRPhenomD aligned-spin waveform model~\cite{Khan:2015jqa} is used in this analysis, and we sample
over the resulting binary black hole parameter space. The parameter definitions and the priors used in
the analysis are as listed in~\cite{Prathaban:2025qgg}. We do not include any additional
parameters in the analysis to account for calibration uncertainties, which enables a direct comparison with~\cite{wong2023fastgravitationalwaveparameter, Polanska:2024zpn}.



\begin{figure}[ht!]
    \centering
    \begin{subfigure}[b]{0.48\linewidth} 
        \includegraphics[width=\linewidth]{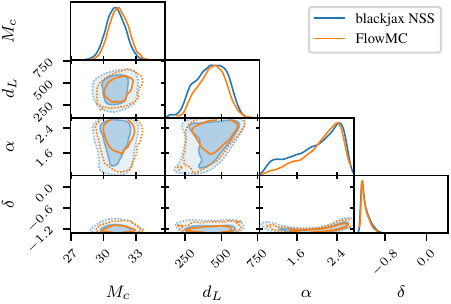}
        \caption{Comparison between the GPU-based \texttt{blackjax} nested sampler and FlowMC for the posterior on the chirp mass, luminosity distance, and sky position in the GW150914 event.}
        \label{fig:reduced_corner}
    \end{subfigure}
    \hfill
    \begin{subtable}[b]{0.48\linewidth} 
        \centering
        \setlength{\tabcolsep}{4pt} 
        \begin{tabular*}{\linewidth}{@{\extracolsep{\fill}} lcr}
            \toprule
            Algorithm & Runtime (s) & ESS \\
            \midrule
            \texttt{blackjax} nss & 207 & 17490 (7599)  \\ 
            FlowMC & 742 & 13633\\
            bilby$*$ & $10^4$ & 5130\\
            \bottomrule
        \end{tabular*}
        \caption{Runtime for sampling GW150914, where $*$ indicates values taken from~\cite{wong2023fastgravitationalwaveparameter}, the bracket ESS values refer to equal weight samples.}
        \label{tab:sample-table}
        \vspace{0.5in}
    \end{subtable}
    \caption{Runtime and posterior inference on GW150914.}
    \label{fig:combined_results}
\end{figure}

\begin{figure}
    \centering
    \includegraphics[width=0.49\linewidth]{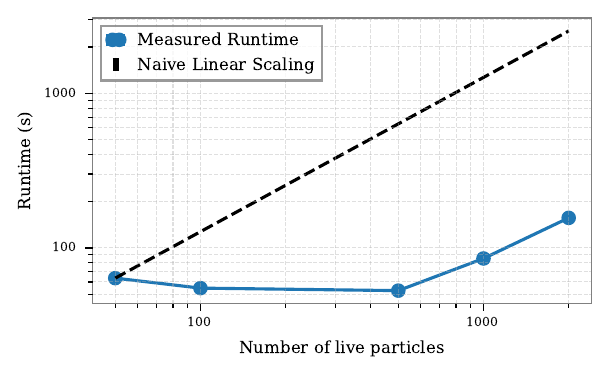}
    \hfill
    \includegraphics[width=0.49\linewidth]{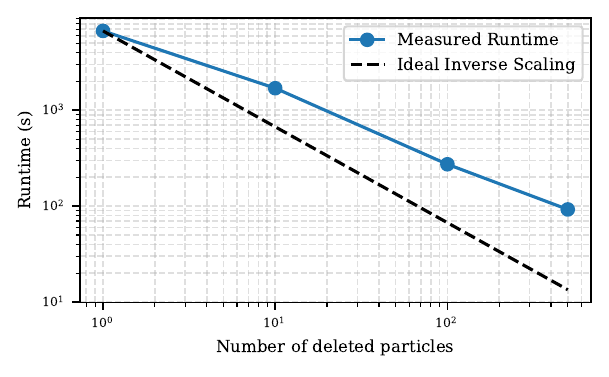}
    \caption{Runtime scaling for nested sampling inference with a heterodyned likelihood on the GW150914 event. Left shows the runtime scaling with a number of deleted particles fixed to half the number of live points, the naive linear scaling expected if the algorithm is not parallelised is shown as a dashed line. Right shows the runtime scaling for a fixed number of 1000 live points as the number of deleted particles is scaled, this time the best case of perfect parallelism is shown as the dashed line.}
    \label{fig:scaling}
\end{figure}

We present the runtimes and effective sample sizes (ESS) of the
resulting posterior samples in Table \ref{tab:sample-table}. The
\texttt{blackjax} NSS implementation achieves a runtime of 207
seconds, demonstrating a significant speedup compared to the
CPU-based implementation of \texttt{bilby} (runtime taken
from \cite{wong2023fastgravitationalwaveparameter}), while also
converging almost 3 times as fast as FlowMC. Further, we find that
\texttt{blackjax} NSS achieves a substantially higher ESS per second than both \texttt{bilby} (\texttt{dynesty}) and FlowMC. We evaluate the ESS of FlowMC via the standard measure implemented in the \texttt{arviz} package~\cite{arviz_2019}, and compute the ESS of \texttt{blackjax} nested sampling chain using the \emph{kish} measure as implemented in the \texttt{anesthetic} package~\cite{anesthetic}.
This indicates that the \texttt{blackjax} implementation is more
efficient at exploring the posterior distribution per unit of
computational time. Whilst some of the computational cost of FlowMC is amortized in the global density proposal, affording increased efficiency asymptotically, similar schemes have been proposed for nested sampling that could greatly enhance this method in a similar manner~\cite{Prathaban:2024rmu}. The marginalized posteriors are plotted in~\cref{fig:reduced_corner} for a reduced set of the full parameter space that is explored, we note that both algorithms have converged to very similar distributions. Performing some ablations of parameters controlling the runtime suggests that these are conservative, but reliable algorithm hyperparameters for both algorithms on this task.
This demonstrates our GPU-accelerated
nested sampling pipeline as a viable method for robust and efficient
GW parameter estimation, and slice sampling can provide a robust alternative to the standard parallel-walk. We demonstrate the parallel nature of the algorithm in this regime by studying the total runtime on the same parameter estimation problem whilst varying two hyperparameters of the algorithm in \cref{fig:scaling}. We demonstrate that by increasing the size of the live population, or by increasing the deleted fraction of the population, significant gains in runtime are possible. This scaling analysis is run on a single NVIDIA L4 GPU.

Ultimately we chose to focus on parameter estimation as the primary task in this work, but importantly the \texttt{blackjax} nested sampling implementation is itself a classical nested sampling algorithm, and thus can be used to compute the Bayesian evidence for model comparison. Validating the accuracy of this estimation, in light of ML assisted techniques~\cite{Polanska:2024zpn}, alongside exploration of more advanced waveform models and likelihoods, is a highlighted area for future work.


\section{Conclusions}\label{sec:conclusion}

In this work we have demonstrated the application of our GPU-accelerated
nested sampling implementation to the analysis of real gravitational
wave data from the GW150914 event. Our key results, presented in Table
\ref{tab:sample-table} and Figure \ref{fig:reduced_corner}, show a
significant improvement in computational efficiency compared to
established CPU-based methods using \texttt{bilby}, achieving runtime speedups
by two orders of magnitude while maintaining a high Effective Sample Size
(ESS). We draw direct comparison to a similarly GPU-accelerated likelihood based MCMC sampler, FlowMC~\cite{Wong:2022xvh}, and find that the \texttt{blackjax} nested sampling implementation converges in a comparable runtime and yields a higher ESS per second. This is despite limited tuning of the slice sampling kernel which we expect to improve these results even further.
Looking forwards, whilst not explored here, our nested sampling approach also directly yields reliable evidence estimates with informative error bars for no extra computational cost, simplifying the parameter estimation and model
comparison process. These results underscore the potential of our
method to accelerate the analysis of gravitational wave signals,
paving the way for more efficient and comprehensive investigations
of future gravitational wave events.

The impressive parallelism exhibited by GPU nested sampling will be a crucial focus for the broader field of astrophysical inference going forward. As available computational resources shift further towards GPUs, algorithms that can exploit the parallelism opportunities of these devices will be essential. Nested Sampling is already well established as a strong baseline for Bayesian inference across the field, and this work demonstrates that nested sampling is not just a legacy baseline, but a powerful and efficient tool for the future.

\section*{Acknowledgements}
MP is supported by the Harding Distinguished Postgraduate Scholars Programme (HDPSP). JA is supported by a fellowship from the Kavli Foundation. The authors were supported by the research environment and infrastructure of the Handley Lab at the University of Cambridge. We thank the \texttt{jim} and \texttt{ripple} authors for the public codes that were influential to this work.

\bibliography{SciPost_Example_BiBTeX_File.bib}

\end{document}